\documentclass[useAMS,usenatbib,usegraphicx]{mn2e}
\usepackage{times}
\usepackage{booktabs}

\title[Evolution of the extinction curves]{Evolution of the extinction
curves in galaxies}
\author[Asano et al.]{Ryosuke S. Asano$^{1}$\thanks{E-mail:
asano.ryosuke@g.mbox.nagoya-u.ac.jp}\thanks{Fellow of the Japan Society for the Promotion of Science (JSPS).},
Tsutomu T. Takeuchi$^{1}$,
Hiroyuki Hirashita$^{2}$
and Takaya Nozawa$^{3}$ 
\\
$^{1}$Department of Particle and Astrophysical Science, Nagoya University, Furo-cho, Chikusa-ku, Nagoya 464-8602, Japan\\
$^{2}$Institute of Astronomy and Astrophysics, Academia Sinica, P.\ O.\
Box 23-141, Taipei 10617, Taiwan\\
$^{3}$Kavli Institute for the Physics and Mathematics of the Universe (WPI),
University of Tokyo, Kashiwa,
Chiba 277-8583, Japan}
\begin{document}
 
\date{Accepted **. Received **; in original form **}

\pagerange{\pageref{firstpage}--\pageref{lastpage}} \pubyear{2014}

\maketitle
   
\label{firstpage}

\begin{abstract}
We investigate the evolution of extinction curves in galaxies based on
 our evolution model of grain size distribution.
In this model, we considered various processes: dust formation by
 SNe~II and AGB stars, dust destruction by SN shocks in the ISM, metal
 accretion onto the surface of grains (referred to as grain growth), shattering
 and coagulation.
We find that the extinction curve is flat in the earliest stage of
 galaxy evolution. 
As the galaxy is enriched with dust, shattering becomes effective
to produce a large abundance of small grains ($a \la
 0.01\;\mu$m).
 Then, grain growth becomes effective at small grain radii, forming a bump at $a
 \sim 10^{-3}${--}$10^{-2}\;\mu$m on the grain size distribution.
Consequently, the extinction curve at ultraviolet (UV) wavelengths becomes steep,
and a bump at $1/\lambda \sim 4.5\;\mu{\rm
 m}^{-1}\;(\lambda: \mbox{wavelength})$ on the extinction curve becomes prominent.
Once coagulation becomes effective, the extinction curves become flatter, 
but the UV extinction remains overproduced when
 compared with the Milky Way extinction curve.
This discrepancy can be resolved by introducing a stronger contribution of coagulation.
Consequently, an interplay between shattering and coagulation could be important
 to reproduce the Milky Way extinction curve.
\end{abstract}


\section{Introduction}
\label{sec:intro}
Dust grains are one of the fundamental ingredients for understanding the
formation and evolution of galaxies.
The surface of dust grains is the main site for the formation of
hydrogen molecules \citep[e.g.,][]{cazaux04}, which act as an
effective coolant in the low-metallicity condition \citep[e.g.,][]{hirashita02,cazaux09}.
Dust is also an important coolant in star formation, inducing a
fragmentation into low-mass stars \citep{omukai,schneider}.
Thus, dust grains are strongly related to the star formation in galaxies.
Also, dust grains govern the scattering and absorption (i.e., extinction) of stellar light,
in particular, at short wavelengths like ultraviolet (UV), and re-emit in infrared (IR).
Consequently, dust grains affect the spectral energy distribution (SED)
of galaxies significantly \citep[e.g.,][]{takagi}.

The extinction curve, which represents the wavelength dependence of
dust extinction, is used to relate the intrinsic stellar SED
with the observed SED affected by dust extinction.
Thus, the extinction curve is the fundamental tool in interpreting the
observational SED of galaxies.
Since the extinction curve depends strongly on the physical
and optical properties of dust grains (grain size, dust components,
etc.) [e.g., \citealt{MRN} (hereafter MRN); \citealt{WD01,nozawa13}], it is important to
understand those properties.

The mean extinction curve of the Milky Way (hereafter MW extinction curve) is observationally
well investigated \citep[e.g.,][]{fitzpatrick}, and is widely adopted
as a template extinction curve in various studies \citep[e.g.,][]{buat,matsuoka,kobayashi13,kruhler}.
The MW extinction curve has a bump at $2175$~\AA~which is
thought to be generated by small carbonaceous grains and/or
polycyclic aromatic hydrocarbons (PAHs) \citep[e.g.,][]{barbaro01,draine09a}, and shows
a steep rise to the far-UV wavelength (called UV slope).
By fitting the MW extinction curve, MRN derived the grain size distribution in the
Milky Way, $f(a){\rm d}a \propto a^{-3.5}{\rm d}a$ with $a = 0.005$--$0.25\;\mu$m,
where $a$ is the grain radius and $f(a){\rm d}a$ is the number density
of grains in size interval [$a, a + {\rm d}a$].
Furthermore, \citet{WD01} performed a detailed fit to the MW extinction
curve, finding that the size distributions of
carbonaceous and silicate dust grains are quite different from each other
unlike MRN.
The extinction curves depend on the line of sight, and
\citet{cardelli} suggested that the variation of the extinction curves are described by adopting the 
parameter $R_V \equiv A_V/E(B - V)$, where $A_V$ is the magnitude of the
extinction in the $V$ band and $E(B - V)$ is the reddening ($A_B - A_V$,
where $A_B$ is the extinction in the $B$ band).
Recently, \citet{nozawa13} investigated the possible variety of
dust properties based on the diversity of the extinction curves observed
in the Milky Way. They found that the power-law index and maximum radius of the
grain size distribution are tightly constrained to be $-3.5 \pm 0.2$ and
$0.25 \pm 0.05\;\mu$m, respectively. 
\citet{pei} extended the graphite-silicate grain model which can fit the
MW extinction curve (MRN) to the Large and Small Magellanic Clouds
(LMC and SMC), and found that the extinction curves in these galaxies
can be fitted with the MRN grain size distribution by adjusting only the
relative contribution of graphite and silicate.

Many studies have shown that high-$z$ galaxies
have different extinction curves from nearby galaxies
\citep[e.g.,][]{maiolino,liang,gallerani,hjorth}. 
The extinction curve of the quasar SDSS104845.05$+$463718.3 (hereafter SDSS1048$+$4637) at redshift
$z = 6.2$ shows the lack of the
$2175$~\AA~bump, and is relatively flat at $\lambda \ga 1700$~\AA~and rising
toward shorter wavelengths at $\lambda \la 1700$~\AA.
\citet{maiolino}, by using the model in \citet{todini}, showed that the
extinction curve is consistent with the dust formation in 
Type~II supernovae (SNe~II).
Because of their short lifetime (typically $10^{6-7}$ yr), SNe~II
are thought to be the origin of dust in high-$z$ Universe.
On the other hand, because of the long lifetime of progenitors,
asymptotic giant branch (AGB) stars can be dominant
sources of dust in galaxies at age $t > 1$~Gyr \citep[but see][]{valiante09}.
Furthermore, \citet{hirashita10b} examined the extinction curves in
starburst galaxies taking into account not only dust grains produced by
SNe~II but also the effect of shattering (grain{--}grain collision) in
the warm ionized medium (WIM).
They showed that the shattering can lead to the steepness of the
extinction curve at UV wavelengths, and indicated that shattering may
occur effectively in SDSS1048$+$4637.
\citet{liang} showed that the extinction curves of high-$z$
gamma-ray bursts (GRBs) are different from those of the Milky Way and LMC.
Among them, one at $z = 6.3$ appears to have the
$2175$~\AA~feature, indicating a difference from SDSS1048$+$4637.
In order to reveal the origin of the differences in the extinction
curves among galaxies at high and low-$z$ Universe,
it is necessary to clarify the processes that govern the evolution of
dust grains in galaxies.

An MRN-like power-law grain size distribution can be realized
if the grains are processed by the grain{--}grain
collisions (shattering and coagulation) \citep[e.g.,][]{tanaka,kobayashi10}.
Thus, it is probable that the grain{--}grain
collisions are an important process in the Milky Way and perhaps in
nearby galaxies in general.
In addition, if the metallicity in galaxies is larger than a certain
value, the accretion of gas-phase metals on the surface of pre-existing grains (referred
to as `grain growth' in this paper) occurs effectively
\citep{inoue11,asano13a}.
Since grain growth has a potential to change the grain size
distribution [e.g., \citealt{hirashita11,asano13b} (hereafter A13)], 
the shape of extinction curves may change by grain growth.
In fact, \citet{hirashita12} showed that the UV slope on extinction
curves becomes steeper by grain growth if the grain size distribution is
initially similar to the MRN size distribution.

Various kinds of processes have different effects on different grain sizes.
Grains ejected by SNe~II into the ISM are
relatively large ($a \ga 0.01\;\mu$m) due to the destruction of small
grains by the sputtering in reverse shocks
\citep[e.g.,][]{bianchi,nozawa07}.
Grains produced in AGB stars may have typical radii $\sim 0.1\;\mu$m \citep[e.g.,][]{winters,ventura12a,
ventura12b,dicriscienzo} and could be described by a log-normal distribution with a
peak at $a \sim 0.1\;\mu$m \citep{yasuda}.
Further, small grains ($a < 0.01\;\mu$m) are efficiently destroyed by
sputtering in interstellar shocks
driven by SNe \citep{nozawa06}.
When grain growth occurs in the interstellar medium (ISM), smaller
grains grow more efficiently (e.g., \citealt{hirashita11}; A13)
because the timescale of grain growth is proportional to
the volume{-}to{-}surface ratio of a dust grain.
In the diffuse ISM, shattering can occur effectively if grains
are dynamically coupled with
magnetized interstellar turbulence \citep[e.g.,][]{yan,hirashita09b},
in particular, large grains ($a \ga 0.1\;\mu$m) acquire larger
velocity dispersions than the shattering threshold velocities.
Shattering occurs also in SN shocks \citep[e.g.,][]{jones96}.
In dense and cold regions, coagulation of small grains can occur \citep[e.g.,][]{hirashita09b,ormel};
consequently, the grain size distribution shifts towards larger sizes
\citep[e.g.,][]{hirashita13b}.
After all, the above various processes affecting grain size distribution
(referred to as `dust processes' in this paper) occur in a way dependent on
the metallicity, total dust amount, and grain size distribution, and
could be interrelated.
Thus, it is mandatory to construct a model by taking into account all
dust processes in a unified framework.

There are some studies on the evolution of the grain size
distribution in galaxies \citep[e.g.,][]{liffman,odonnell,hirashita10b,yamasawa}.
However, they did not consider all the dust processes to simplify their models.
Recently, A13 have discussed the evolution of the grain size
distribution, taking into account all the dust processes based on chemical
evolution of galaxies.
A13 showed that the grain size distribution drastically changes with
the galactic age because the dominant dust process changes (see Section $2$).
In view of the discussion in A13, it is expected that the extinction curve 
also changes with the galactic age due to the change of the dominant dust
processes.
Therefore, in this paper, we examine the evolution of extinction
curves in galaxies using the dust evolution model developed by A13, and check whether
we can reproduce the MW extinction curve.

This paper is organized as follows.
First, we briefly review the dust evolution model constructed by
A13 and explain the theoretical treatment of the extinction curve in
Section \ref{sec:model}.
In Section \ref{sec:result}, we show the contributions of
various dust processes to the extinction curve.
We discuss how we can reproduce the MW extinction curve and the
contribution of different grain species in Section \ref{sec:discuss}.
We present the conclusions of this paper in Section \ref{sec:conclusion}.

\section{Model}
\label{sec:model}

In this section, we first review our dust evolution model for calculating the
evolution of the grain size distribution in a galaxy (A13).
Then, we explain the method of calculating the extinction at wavelength
$\lambda$, $A_{\lambda}$ (in units of magnitude) based on the grain size
distribution calculated.

\subsection{Dust evolution model}
\label{subsec:dustmodel}

We briefly introduce the model constructed by A13 and their results.
A13 investigated the evolution of the grain size
distribution taking into account the dust formation by SNe~II and AGB
stars, dust destruction by SN shocks in the ISM, grain growth in the
cold neutral medium (CNM), grain{--}grain collisions (shattering and
coagulation) in the warm neutral medium (WNM) and CNM.
Grain growth in the WNM was not considered because grain growth
is more efficient in dense and cold regions \citep[e.g.,][]{liffman,draine09a}.
A13 considered the contribution of the dust processes
in the WNM and CNM by introducing the mass fractions of WNM ($\sim
6000$~K, $0.3\;{\rm cm}^{-3}$) and CNM ($\sim 100$~K, $30\;{\rm
cm}^{-3}$), $\eta_{\rm WNM}$ and $\eta_{\rm CNM}$, respectively.
The sum of $\eta_{\rm WNM}$ and $\eta_{\rm CNM}$ was assumed to be
unity in A13 since an equilibrium state of two thermally
stable phases (WNM and CNM) is established in the ISM if we consider
temperatures less than $10^4$~K in the ISM \citep{wolfire}.
The grain velocities in the two ISM phases derived by 
\citet{yan} were adopted to calculate shattering and coagulation.

We assume two dust species, graphite and silicate \citep{dralee} in the
same way as in A13.
Although A13 considered a variety of dust species \citep[C, Si,
$\mbox{SiO}_2$, SiC, Fe, FeS, $\mbox{Al}_2\mbox{O}_3$, MgO,
MgSi$\mbox{O}_3$, $\mbox{Mg}_2\mbox{SiO}_4$ and
$\mbox{Fe}_2\mbox{SiO}_4$;][]{nozawa07,zhukovska} for stellar dust production,
carbonaceous dust and all the other dust species are categorized as
graphite and silicate [we adopt astronomical silicate,
$\mbox{Mg}_{1.1}\mbox{Fe}_{0.9}\mbox{SiO}_4$ \citep{dralee}],
respectively to avoid chemical complexity in grain growth, shattering,
and coagulation. A13 examined the contribution of grain growth, shattering
and coagulation for two dust species separately.
While the complexity of dust species may affect the extinction curves
\citep[e.g.,][]{nozawa13}, the aim of this paper is not a detailed fit
to a specific extinction curve but an investigation of the response of
extinction curves to the evolution of grain size distribution.
Further, it is thought that these two dust species are the main components of dust
grains in the Milky Way \citep{dralee},
and \citet{sofia} suggested that (Fe$+$Mg):Si ratio in dust
grains is close to $2$:$1$.
Thus, the above grain composition is a reasonable approximation.
Other possible dust species are discussed in Section~$4$.
We assume that grains are spherical,
and that shattering/coagulation occurs if the relative velocity of collisional grains is more/less than
the threshold velocity \citep[e.g.,][]{yan,hirashita09b}.
The threshold velocity of shattering, $v_{\rm shat}$, is assumed to
be $1.2$ and $2.7\;{\rm km}\;{\rm s}^{-1}$ for carbonaceous dust and
silicate dust, respectively \citep{jones96}.
For the threshold velocity of coagulation, we calculate in the same way
as \citet{hirashita09b}, and the threshold velocity is about
$10^{-3}${--}$10^{-1}\;{\rm km}\;{\rm s}^{-1}$ depending on the grain
size \citep{chokshi}.
Note that if the radius of the shattered fragments is less than $3$~\AA,
we remove the fragments unlike A13.

For the dust evolution model, A13 assumed that the total baryon mass (the sum of the
stellar mass and the ISM mass in the galaxy) is constant (closed-box
model), and formulated the star formation rate (SFR) by introducing the
star formation timescale, $\tau_{\rm SF}$: $\mbox{SFR}(t) = M_{\rm
ISM}(t)/\tau_{\rm SF}$, where $M_{\rm ISM}$ is the ISM mass and $t$ is the galaxy age.

\begin{figure}
\centering\includegraphics[width=0.4\textwidth]{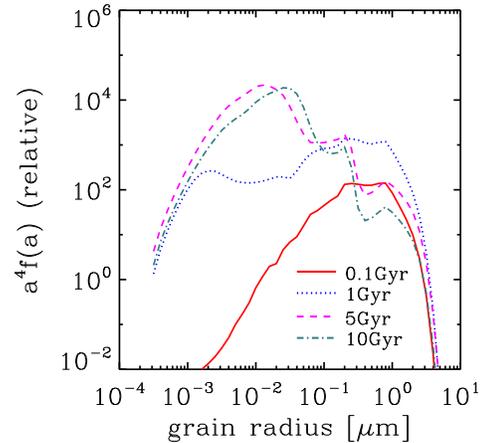}
\caption{Example of the evolution of the grain size distribution.
Solid, dotted, dashed, and dot{--}dashed lines represent
 the cases at $t = 0.1, 1, 5, 10$~Gyr, respectively, with $\tau_{\rm
 SF} = 5$~Gyr. The mass fractions, $\eta_{\rm WNM}$ and $\eta_{\rm
 CNM}$, are set to be $0.5$.
}
\label{fig:gsd}
\end{figure}
Figure~\ref{fig:gsd} shows the evolution of the grain size
distribution (the sum of silicate and carbonaceous dust is shown) with
all dust processes considered for $\tau_{\rm SF} = 5$~Gyr.
We find that while the grain size distribution is dominated by
large grains ($a \ga 0.1\;\mu$m) produced by stars at $< 0.1$~Gyr, as
the galaxy evolution proceeds, the grain size distribution begins to
be regulated by the processes in the ISM.
In particular, once shattering occurs effectively, a large amount of small
grains ($a \la 0.01\;\mu$m) are produced by the fragmentation due to collisions between large grains. 
The effect of shattering is seen in the increase of small grains at $0.1${--}$1$~Gyr.
Due to a large amount of small grains, grain growth occurs
effectively because the surface-to-volume ratio
of smaller grains is larger than that of larger grains.
Consequently, the bump at $a \sim 0.01\;\mu$m emerges at
$1${--}$10$~Gyr.
Smaller grains can acquire lower velocity dispersions since they are coupled with smaller scale turbulence \citep[e.g.,][]{yan}.
Therefore, after small grains are enhanced, the coagulation mainly
occurs by collisions between small grains whose velocity dispersions are
smaller than the coagulation threshold.
The shift of the bump position from $1$ to $10$~Gyr is due to coagulation.
For further details and parameter dependences of grain size
distribution, see A13.

\subsection{Extinction curve}

Extinction curves are powerful tools to examine the
dust properties in galaxies.
In order to analyze extinction curves, the optical constants for
each dust species are necessary.
In this paper, we adopt the optical constants derived by \citet{dralee}
to calculate the grain extinction cross section normalized to the
geometrical cross section $\pi a^2$ as a function of
wavelength and grain radius, $Q_{\rm ext, X}(\lambda,a)$, where the
subscript X represents grain species (X = carbonaceous dust or silicate dust) and $\lambda$ is the wavelength.

The optical depth of dust species X at a given wavelength $\lambda$,
$\tau_{{\rm X}, \lambda}$, is defined as
\begin{equation}
\tau_{{\rm X}, \lambda} = \int^{\infty}_{0} \pi a^2 C Q_{\rm ext,
 X}(\lambda,a) f_{\rm X}(a) {\rm d}a,
\end{equation}
where $C$ is a normalization constant and $f_{\rm X}(a)$ is defined so that
$f_{\rm X} {\rm d}a$ is the number density of species X with radii in the range $[a, a+{\rm d}a]$.
The extinction in units of magnitude is proportional to the optical
depth, and is expressed as
\begin{equation}
A_{{\rm X}, \lambda} = 1.086 \tau_{{\rm X}, \lambda},
\end{equation}
where $A_{{\rm X}, \lambda}$ is the extinction of dust species X in units of
magnitude at wavelength $\lambda$.
The total extinction in units of magnitude, $A_{\lambda}$, is expressed
as
\begin{equation}
A_{\lambda} = \sum_{\rm X} A_{{\rm X}, \lambda}.
\end{equation}
In this work, we consider the extinction curve normalized to the
$V$ band value, $A_{\lambda}/A_{\lambda_V}$,
so $C$ and factor $1.086$ cancel out.

\section{Results}
\label{sec:result}

In this section, we show the effects of each dust process on the
extinction curve; we add the following processes one by one, the dust formation by SNe~II and AGB
stars, dust destruction by SN shocks in the ISM, grain growth in the
ISM, shattering and coagulation in the ISM.
The loss of dust by astration is always included, although it does not
affect the shape of extinction curve.
We adopt star formation timescale $\tau_{\rm SF} = 5$~Gyr, and the mass fraction of the WNM
and CNM $\eta_{\rm WNM} = \eta_{\rm CNM} = 0.5$ unless otherwise stated.

\subsection{Dust formation by SNe~II and AGB stars}

\begin{figure}
\centering\includegraphics[width=0.4\textwidth]{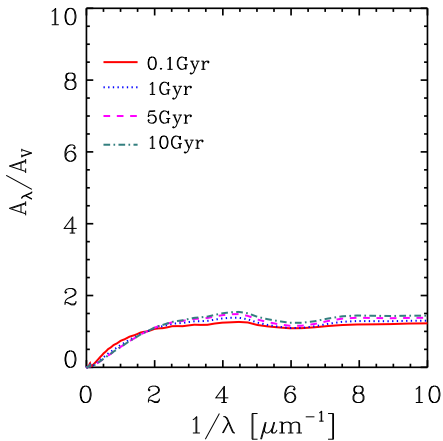}
\caption{Evolution of extinction curves with the dust formation by
 SNe~II and AGB stars. 
Solid, dotted, dashed, and dot{--}dashed lines represent
 the cases at $t = 0.1, 1, 5, 10$~Gyr, respectively, with $\tau_{\rm
 SF} = 5$~Gyr. 
}
\label{fig:staronly}
\end{figure}

In Fig.~\ref{fig:staronly}, we show the evolution of the extinction
curves with the dust formation by SNe~II and AGB stars.
From Fig.~\ref{fig:staronly}, we observe that the extinction curves
are flat throughout any galactic ages, and do not
change significantly with time.
This is because the size distribution of grains produced by SNe~II and
AGB stars is dominated by large grains ($a \ga 0.1\;\mu$m)
\citep{nozawa07,yasuda} and does not change with
galactic ages considerably (Fig.~1 of A13).

A13 showed that although the size distribution of grains produced by
SNe~II depends on the hydrogen number density of the ISM surrounding the
SNe~II, $n_{\rm SN}$ (we adopt $n_{\rm SN} = 1.0\;{\rm cm}^{-3}$ in this
paper), the tendency that the grain size distribution is dominated by
large grains is unchanged.

\subsection{Dust destruction}

\begin{figure}
\centering\includegraphics[width=0.4\textwidth]{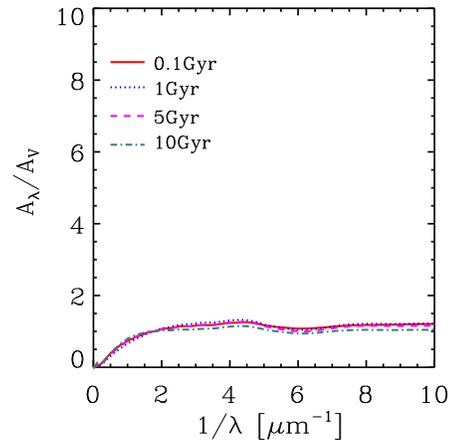}
\caption{Same as Fig.~\ref{fig:staronly} but with the dust destruction by
 SN shocks in the ISM in addition.}
\label{fig:dest}
\end{figure}

In Fig.~\ref{fig:dest}, we show the evolution of extinction curve with the dust destruction by SN shocks in the ISM in
addition to the dust formation by stars.
The timescale on which the dust destruction affects the grain
size distribution in galaxies, $\tau_{\rm SN}$, is
about $\tau_{\rm SN} \sim 0.1 \tau_{\rm SF}$ (A13).
Thus, we can not observe the difference between the extinction curves
with and without dust destruction at $0.1$~Gyr (solid lines in
Fig.~\ref{fig:staronly} and Fig.~\ref{fig:dest}).
The extinction curves with dust destruction
are slightly flatter than those produced by stardust (see
Fig.~\ref{fig:staronly}) at $t \ga 1~$Gyr.
Since smaller grains are more easily destroyed by SN shocks (\citealt{nozawa06,yamasawa}; A13),
the extinction curves become flatter than the case without dust destruction by SN shocks in
the ISM (Fig.~\ref{fig:staronly}).

A13 examined the effect of dust destruction by SN shocks in
the ISM for various $n_{\rm SN}$, and showed that the effect is larger for larger
$n_{\rm SN}$.
Thus, the extinction curve becomes flatter in the case
with larger $n_{\rm SN}$.

\subsection{Grain growth}
\label{subsec:growth}

\begin{figure}
\centering\includegraphics[width=0.4\textwidth]{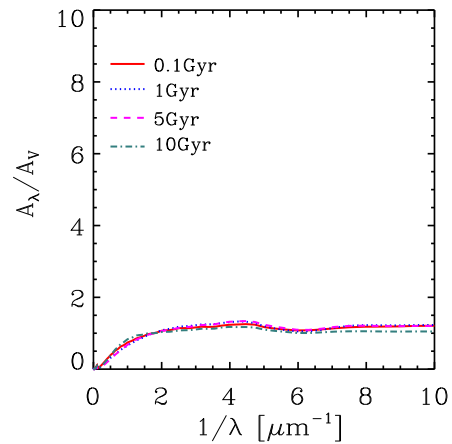}
\caption{Same as Fig.~\ref{fig:dest} but with grain growth in addition.
We adopt $\eta_{\rm CNM} = 0.5$.}
\label{fig:acc}
\end{figure}

Figure \ref{fig:acc} shows the evolution of extinction curves 
with the dust formation by SNe~II and AGB stars, dust destruction by
SN shocks in the ISM and grain growth in the CNM.
We find that the extinction curves are almost the same as in Fig.~\ref{fig:dest}.
In this case, since the total surface area of grains is dominated by
large grains with $a > 0.3\;\mu$m (A13),
the effect of grain growth is prominent at large sizes ($a \ga
0.1\;\mu$m).
Since the grains are already large,
grain growth just keeps the extinction curve flat.

\subsection{Shattering}
\label{subsec:shat}

\begin{figure}
\centering\includegraphics[width=0.4\textwidth]{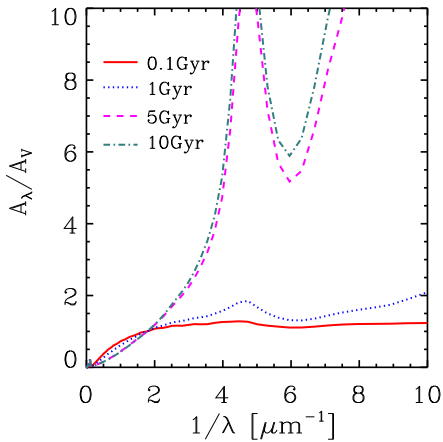}
\caption{Same as Fig.~\ref{fig:acc} but with shattering in addition.
We adopt $\eta_{\rm WNM} = \eta_{\rm CNM} = 0.5$.}
\label{fig:shat}
\end{figure}

We show the evolution of extinction curve with shattering in addition to
all the dust processes considered in Section \ref{subsec:growth}.
From Fig.~\ref{fig:shat}, we observe that the extinction curve is almost
flat until galactic age $t \sim 1$~Gyr;
at $t \la 1$~Gyr, shattering is ineffective due to the
small dust abundance, and the grain size distribution is similar to
the case without shattering (A13).
A13 discussed the timescale on which the grain
size distribution changes due to the processes in the ISM (especially
shattering), and obtained the timescale $\tau_{\rm shat} \sim 1~\left(\tau_{\rm SF}/{\rm
Gyr}\right)^{1/2}$~Gyr from a rough order-of-magnitude estimate (see
their Appendix~B).
Thus, there is only a small difference between the cases with and without
shattering at $t < 1$~Gyr for $\tau_{\rm SF} = 5$~Gyr.
At $t = 5$~Gyr, the extinction curve drastically changes and starts
to have a prominent bump at
$1/\lambda \sim 4.5\;\mu{\rm m}^{-1}$ (the so-called $2175$~\AA~bump)
 and a steep slope toward shorter wavelengths.
This is because the grain size distribution changes considerably due
to the interplay between grain growth and shattering (A13).
A13 showed that once shattering occurs effectively, the number
of small grains increases with the decrease in the number of large grains.
Thus, the total surface area of grains per grain mass becomes large, and
grain growth occurs effectively especially at small grain radii ($\la 0.01\;\mu$m),
forming a bump at $a \sim 0.01\;\mu$m in the grain size distribution.
Consequently, the $2175$~\AA~bump and UV slope on the extinction curve
become larger and steeper, respectively.
In fact, \citet{hirashita12} also showed that the $2175$~\AA~bump and
the UV slope are enhanced by grain growth. 
At $t = 10$~Gyr, although grain growth becomes ineffective due to the
large depletion of heavy elements \citep[e.g.,][]{asano13a},
the abundance of small grains still increases by shattering.
Consequently, the $2175$~\AA~bump and the UV slope become larger and
steeper at $10$~Gyr than at $5$~Gyr.

\subsection{Coagulation}
\label{subsec:coag}

\begin{figure}
\centering\includegraphics[width=0.4\textwidth]{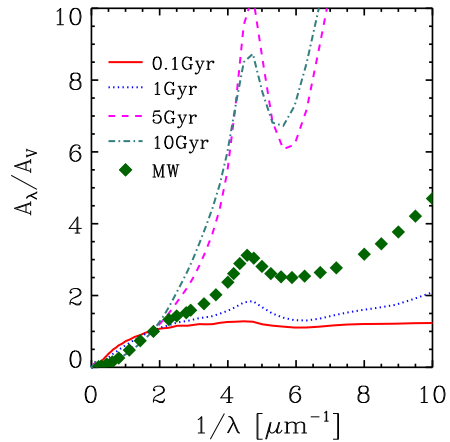}
\caption{Same as Fig.~\ref{fig:shat} but with coagulation in addition.
Filled diamonds represent the extinction curve of the Milky Way \citep{whittet}.
}
\label{fig:coag}
\end{figure}

In Fig.~\ref{fig:coag}, we show the evolution of the extinction curve
with all the dust processes considered in our model.
We find that the extinction curves at
$t = 0.1$ and $1$~Gyr are flat, and that there is little
difference between the extinction curves with and without
coagulation (see Fig.~\ref{fig:shat}).
As mentioned in Section~\ref{subsec:dustmodel}, since coagulation
occurs when the relative velocity of colliding grains is
less than $v_{\rm coag}$ [fiducial $v_{\rm coag} \sim
10^{-3}${--}$10^{-1}\;{\rm km}\;{\rm s}^{-1}$ depending on the grain size
\citep{chokshi}],
coagulation does not occur for large grains ($a \ga 0.1\;\mu$m) whose
velocities are above the coagulation threshold.
At $t = 5$ and $10$~Gyr, the $2175$~\AA~bump is smaller, and the UV slope is flatter than that in the case without
coagulation (Fig.~\ref{fig:shat}). 
A13 showed that the bump at $a \sim 0.01\;\mu$m in the
grain size distribution shifts to larger sizes by coagulation, and 
finally moves to $a \sim 0.03${--}$0.05\;\mu$m at $10$~Gyr.
As the abundance of large grains is
increased by coagulation, the bump at $1/\lambda \sim 4.5\;\mu{\rm m}^{-1}$ and the UV slope
become smaller and flatter, respectively.

For reference, we show the MW extinction curve taken from \citet{whittet}.
From Fig.~\ref{fig:coag}, we find that our calculated extinction curve
at the age of the Milky Way ($\sim 10$~Gyr) is steeper and has a larger $2175$~\AA~bump 
than the extinction curve of the Milky Way.
The main aim of this paper is not to reproduce the MW extinction curve
precisely but to examine the trend of evolution of the extinction curve.
Nevertheless, we will propose and examine some possibilities of reproducing
the MW extinction curve by slightly modifying our model in Section \ref{sec:mw}.

\subsection{Parameter dependence}

So far, we have adopted star formation timescale $\tau_{\rm SF} = 5$~Gyr,
mass fractions of WNM and CNM $\eta_{\rm WNM} = \eta_{\rm CNM} = 0.5$ as
a fiducial case.
Here, we demonstrate how the extinction curve depends on these parameters.

\begin{figure*}
\centering\includegraphics[width=0.4\textwidth]{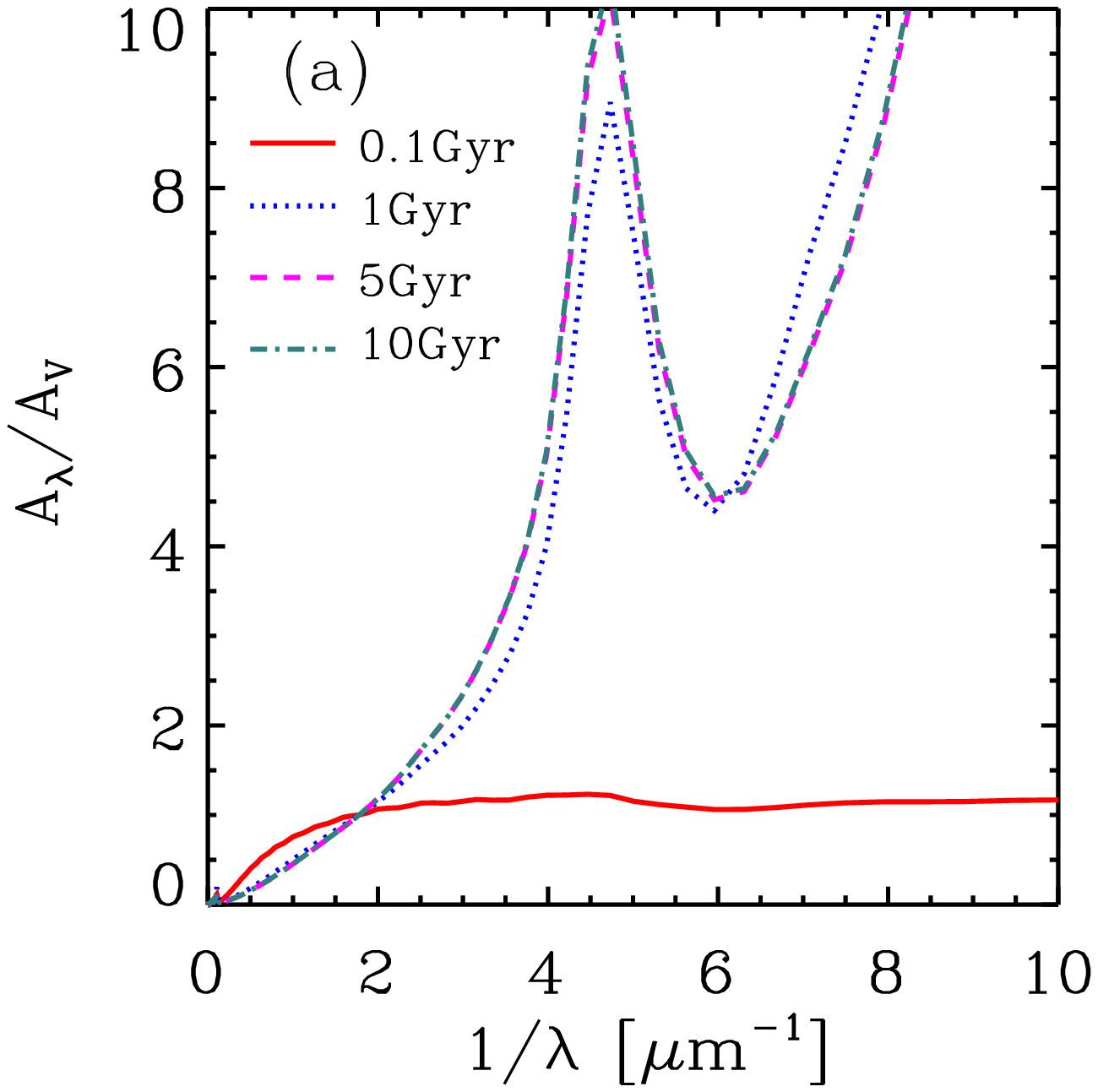}
\includegraphics[width=0.4\textwidth]{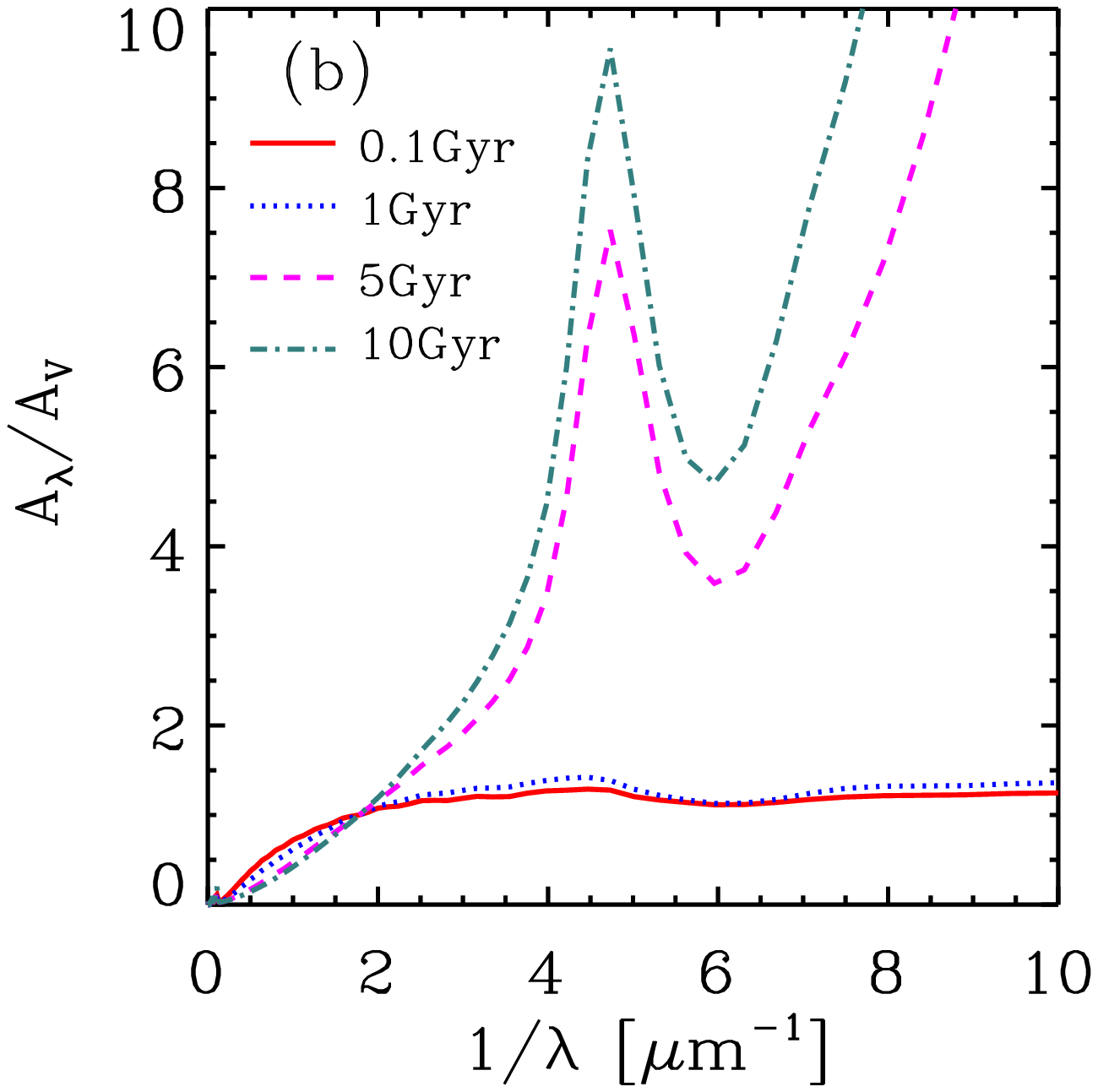}
\includegraphics[width=0.4\textwidth]{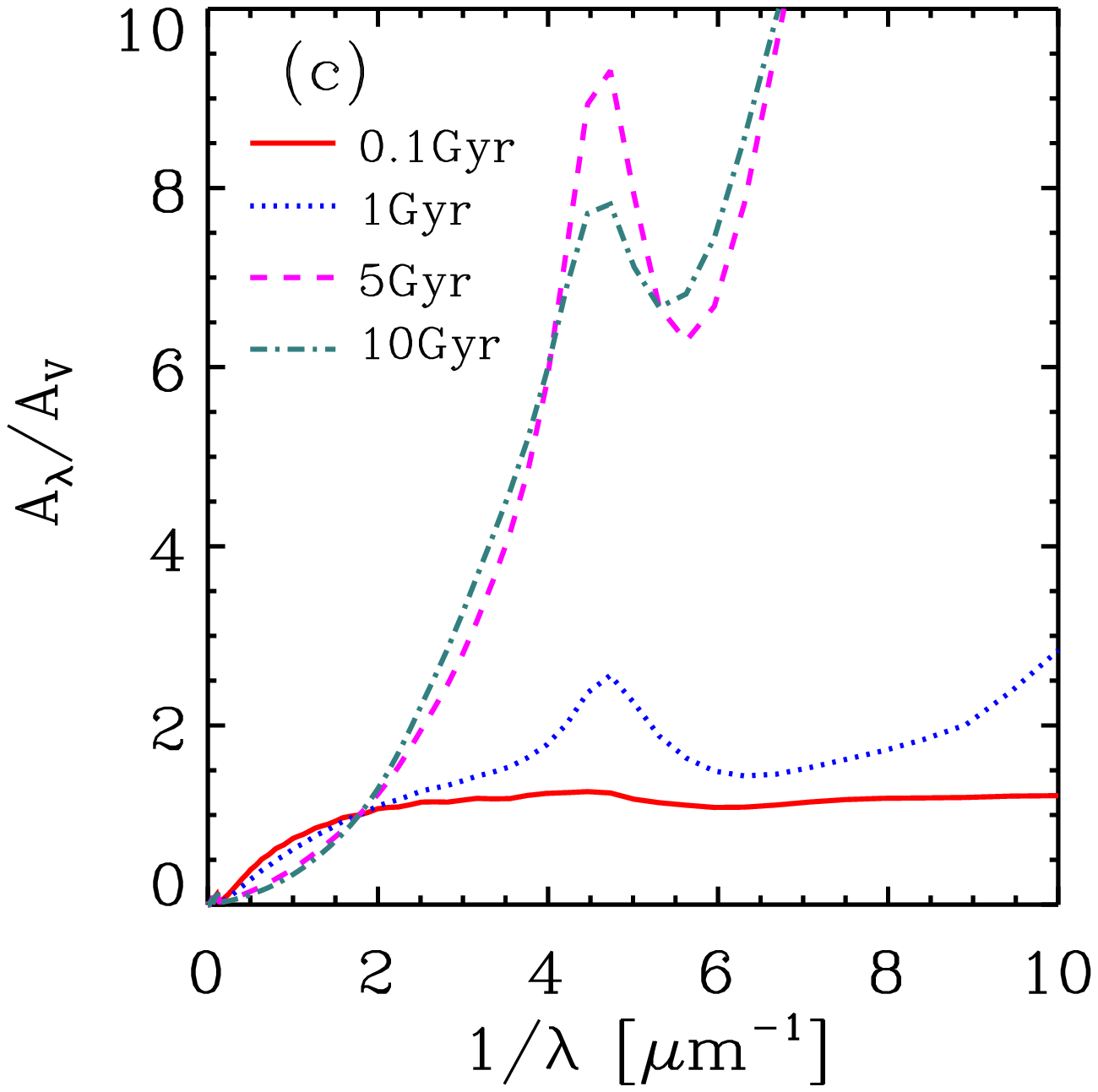}
\includegraphics[width=0.4\textwidth]{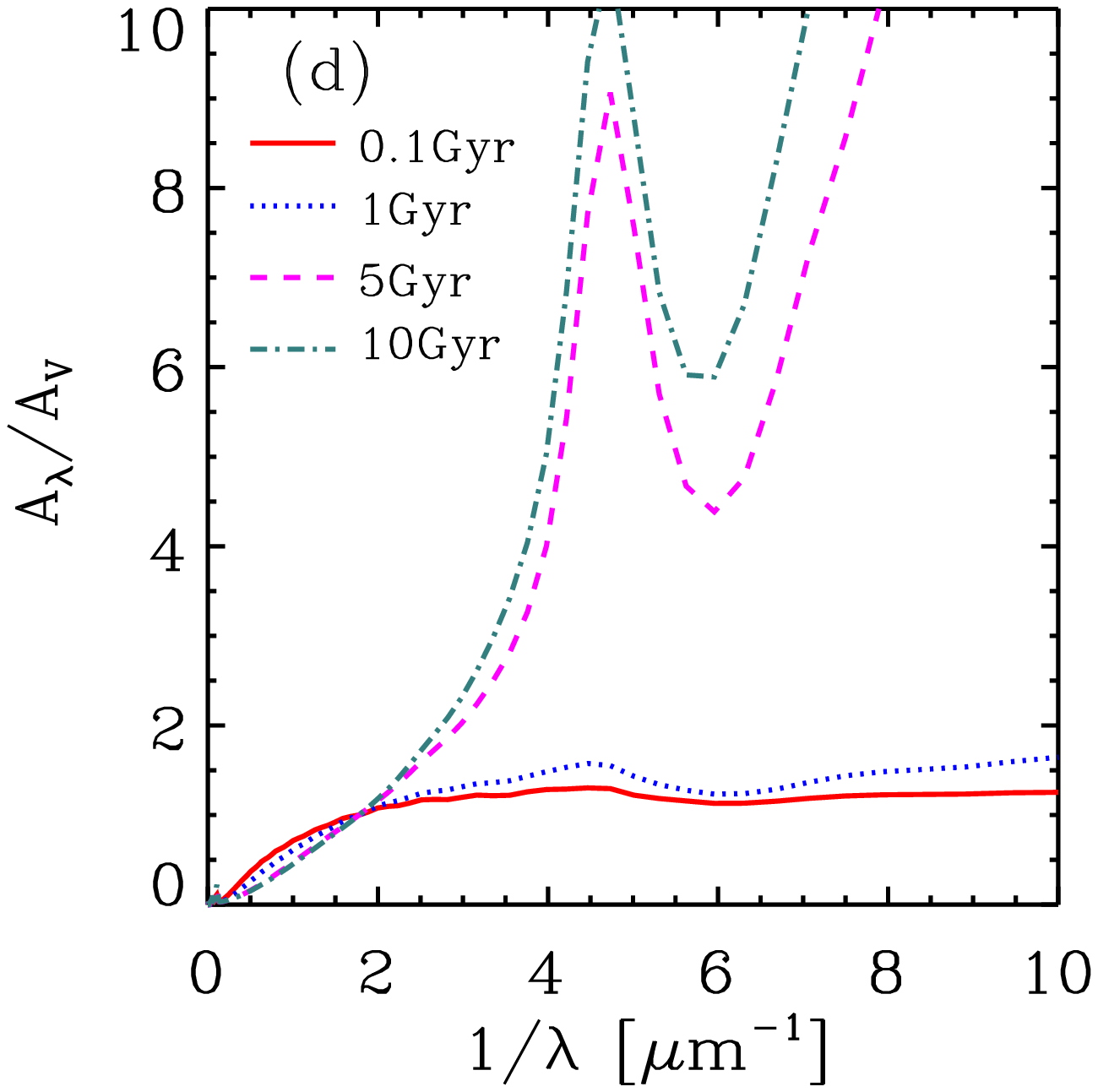}
\caption{Same as Fig.~\ref{fig:coag} but with (a) $\tau_{\rm SF} = 0.5$~Gyr and $\eta_{\rm WNM} =
 \eta_{\rm CNM} = 0.5$, (b) $\tau_{\rm SF} = 50$~Gyr and $\eta_{\rm WNM} =
 \eta_{\rm CNM} = 0.5$, (c) $\tau_{\rm SF} = 5$~Gyr, $\eta_{\rm WNM} =
 0.1$ and $\eta_{\rm CNM} = 0.9$ and (d) $\tau_{\rm SF} = 5$~Gyr,
 $\eta_{\rm WNM} = 0.9$ and $\eta_{\rm CNM} = 0.1$.
}
\label{fig:para}
\end{figure*}

In panels (a) and (b) of Fig.~\ref{fig:para}, we show the time evolution of the extinction
curve in the cases
with $\tau_{\rm SF} = 0.5$ and $50$~Gyr, respectively, for $\eta_{\rm
WNM} = \eta_{\rm CNM} = 0.5$.
Focusing on panel (a), we find that the extinction curve at $1$~Gyr is different
from that in Fig.~\ref{fig:coag} where $\tau_{\rm SF} = 5$~Gyr is adopted.
If $\tau_{\rm SF}$ is short, the total dust amount ejected by stars is
large at younger galactic ages. 
This early increase in dust abundance shortens the timescale on which
shattering becomes effective according to $\tau_{\rm shat} \sim 1 (\tau_{\rm SF}/{\rm
Gyr})^{1/2}$~Gyr (A13; Section \ref{subsec:shat}).
Thus, shattering is already able to produce a large amount of small
grains at $< 1$~Gyr for $\tau_{\rm SF} = 0.5$~Gyr.
Consequently, the extinction curve has the $2175$~\AA~bump and steep UV slope at earlier phases
of galaxy evolution.
We also observe that the extinction curves at $t = 5$ and $10$~Gyr are
almost identical to each other.
This means that the grain size distribution does not change at these ages
for $\tau_{\rm SF} = 0.5$~Gyr.
Since the total dust amount decreases rapidly by astration in the case
with the short
star formation timescale \citep[e.g.,][]{asano13a},
shattering and coagulation become inefficient at those ages.
As a result, the shape of grain size distribution does not change significantly.
On the other hand, panel (b) shows the result for $\tau_{\rm SF} = 50$~Gyr.
We find that the $2175$~\AA~bump continues
to grow even after $5$~Gyr.
This is because the increase of the total dust amount is slower due
to the longer star formation timescale
than the case in panel (a).
Indeed, $\tau_{\rm shat} \sim 7$~Gyr is consistent with the rapid
increase of small grains at $5${--}$10$~Gyr (see also Fig.~9 in A13).
Thus, the timescales of all dust processes we considered
become long (e.g., A13),
and the evolution of the extinction curve slows down.

Panels (c) and (d) of Fig.~\ref{fig:para} show the cases with $(\eta_{\rm
WNM}, \eta_{\rm CNM}) = (0.1, 0.9)$ and $(0.9, 0.1)$, respectively, for $\tau_{\rm SF}
= 5$~Gyr.
Since a large $\eta_{\rm CNM}$ is adopted in panel
(c), the timescale of coagulation is short. Consequently, the $2175$~\AA~bump
becomes small at earlier phases
than the case with a small $\eta_{\rm CNM}$. 
In panel (d) where a small $\eta_{\rm CNM}$ is adopted,
 the timescale of grain growth is longer than that for panel (c).
As mentioned in section \ref{subsec:shat}, the $2175$~\AA~bump
becomes prominent due to grain growth.
Thus, the $2175$~\AA~bump becomes more
prominent at later phases than in the case with a larger $\eta_{\rm CNM}$.

\section{Discussion}
\label{sec:discuss}

\subsection{Reproducing the MW extinction curve}
\label{sec:mw}

In the previous section, we showed the evolution of the extinction curve
in galaxies taking into account various dust processes.
We found that since stellar dust is biased to large grains ($a \ga
0.1\;\mu$m), the extinction curve at the earliest stage of galaxy evolution
is flat.
After $t \sim \tau_{\rm shat} \sim
1~\left(\tau_{\rm SF}/{\rm Gyr}\right)^{1/2}$~Gyr, shattering and grain growth occur effectively, and the
extinction curve becomes steeper and has a larger bump at $1/\lambda
\sim 4.5\;\mu{\rm m}^{-1}$ than that of the MW extinction curve.
After coagulation becomes effective, the bump becomes small.
However, compared with the MW extinction curve,
the calculated extinction curves are too steep and have too large a bump.
The reason for this is 
too abundant small grains, forming the bump at $a \sim 0.01\;\mu$m in the
grain size distribution (Fig.~\ref{fig:gsd}).
Thus, it is crucial to weaken this bump in order to reproduce the MW
extinction curve.
As explained in Section \ref{subsec:shat}, the bump in the grain size
distribution is formed by grain growth.
This means that our models may have overestimated grain growth.
Thus, as the first possibility of reproducing the MW extinction curve,
we examine a model in which only grain growth is tuned off.
This is referred to as (i) no grain growth model.

There is another way of weakening the bump in the grain size
distribution.
In the current model, coagulation takes place only for the grains with
radii less than $\sim 0.05\;\mu$m because
large grains have larger velocities than the coagulation threshold
(Section \ref{subsec:coag}).
However, if the grains can coagulate beyond $0.05\;\mu$m, the bump in
the grain size distribution could be smoothed out.
There are some indications that the grains could coagulate beyond
$0.05\;\mu$m:
\citet{ossenkopf} argued that fluffy grains can be
formed in dense regions by coagulation.
Fluffiness enhances the cross-section in grain{--}grain collision,
raising the coagulation rate.
Fluffy grains can also absorb collision energy, raising the coagulation
threshold \citep[e.g.,][]{ormel}.
Enhancement of coagulation efficiency is also suggested by
\citet{hirashita13b} to explain $\mu$m{--}sized grains in dense molecular
cloud cores.
Thus, to investigate the possibility of strong coagulation, we examine a
model in which we do not apply any coagulation threshold velocity.
This model is referred to as (ii) strong coagulation model.

\begin{figure*}
\centering\includegraphics[width=0.4\textwidth]{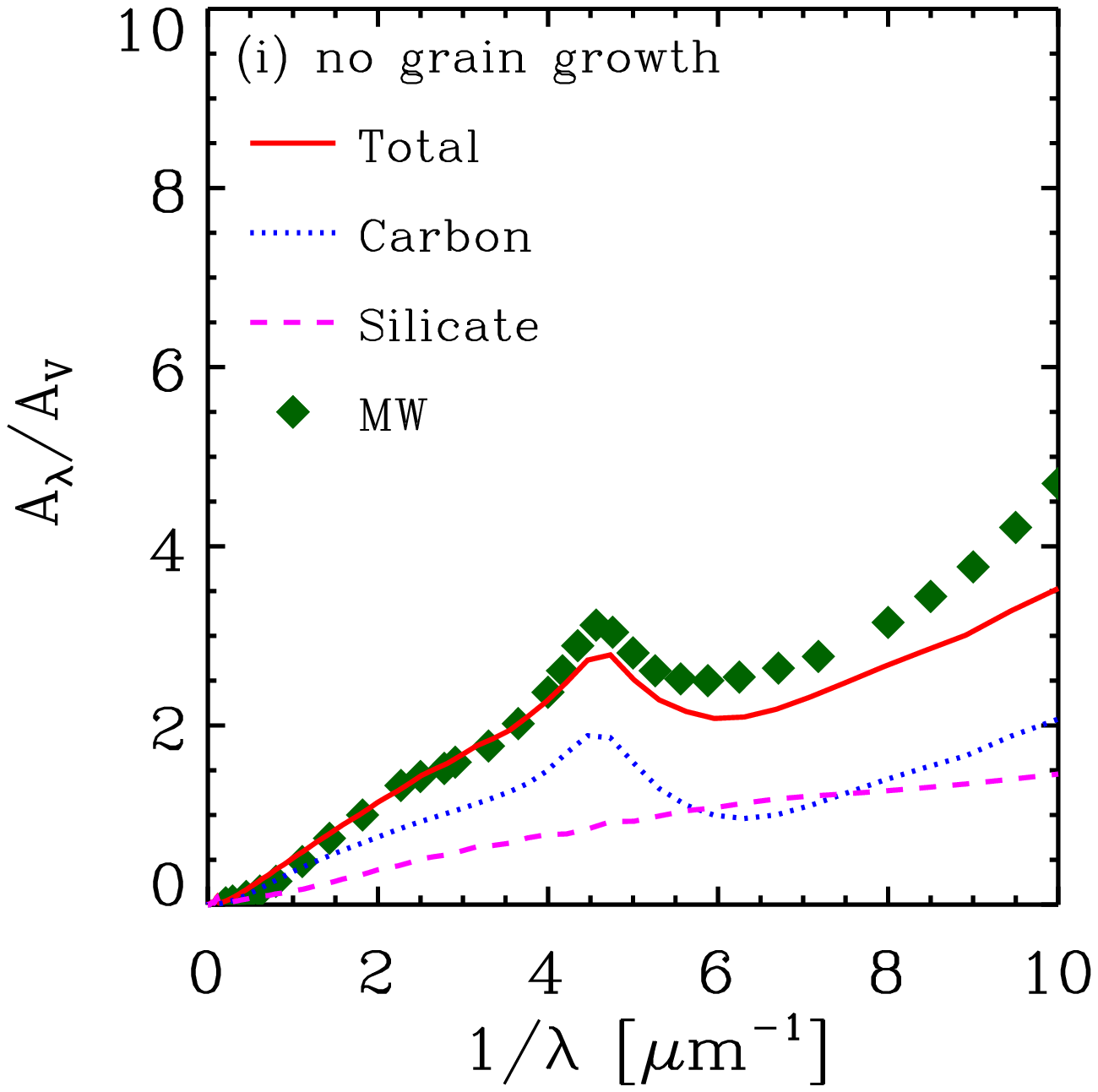}
\includegraphics[width=0.4\textwidth]{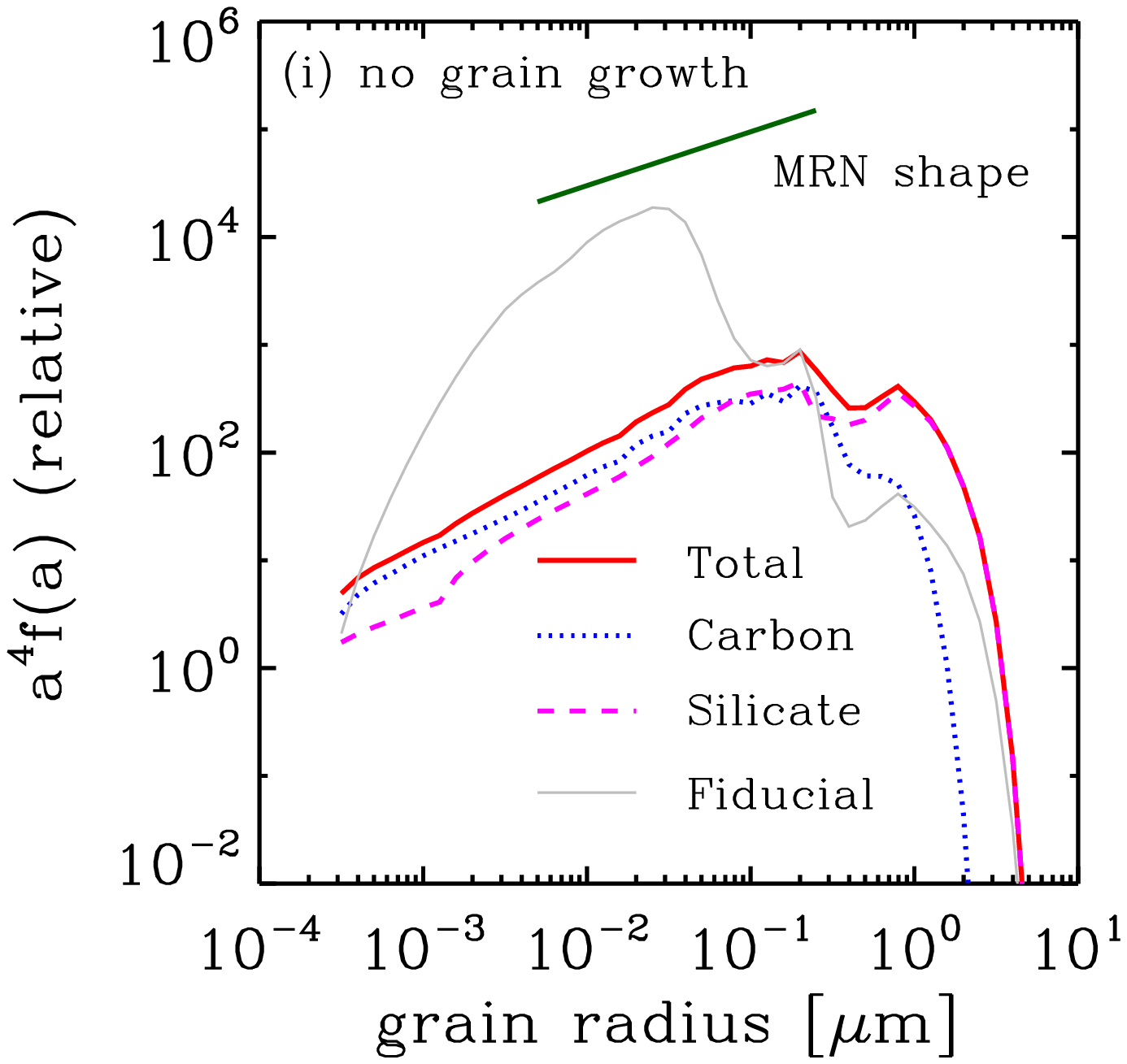}
\includegraphics[width=0.4\textwidth]{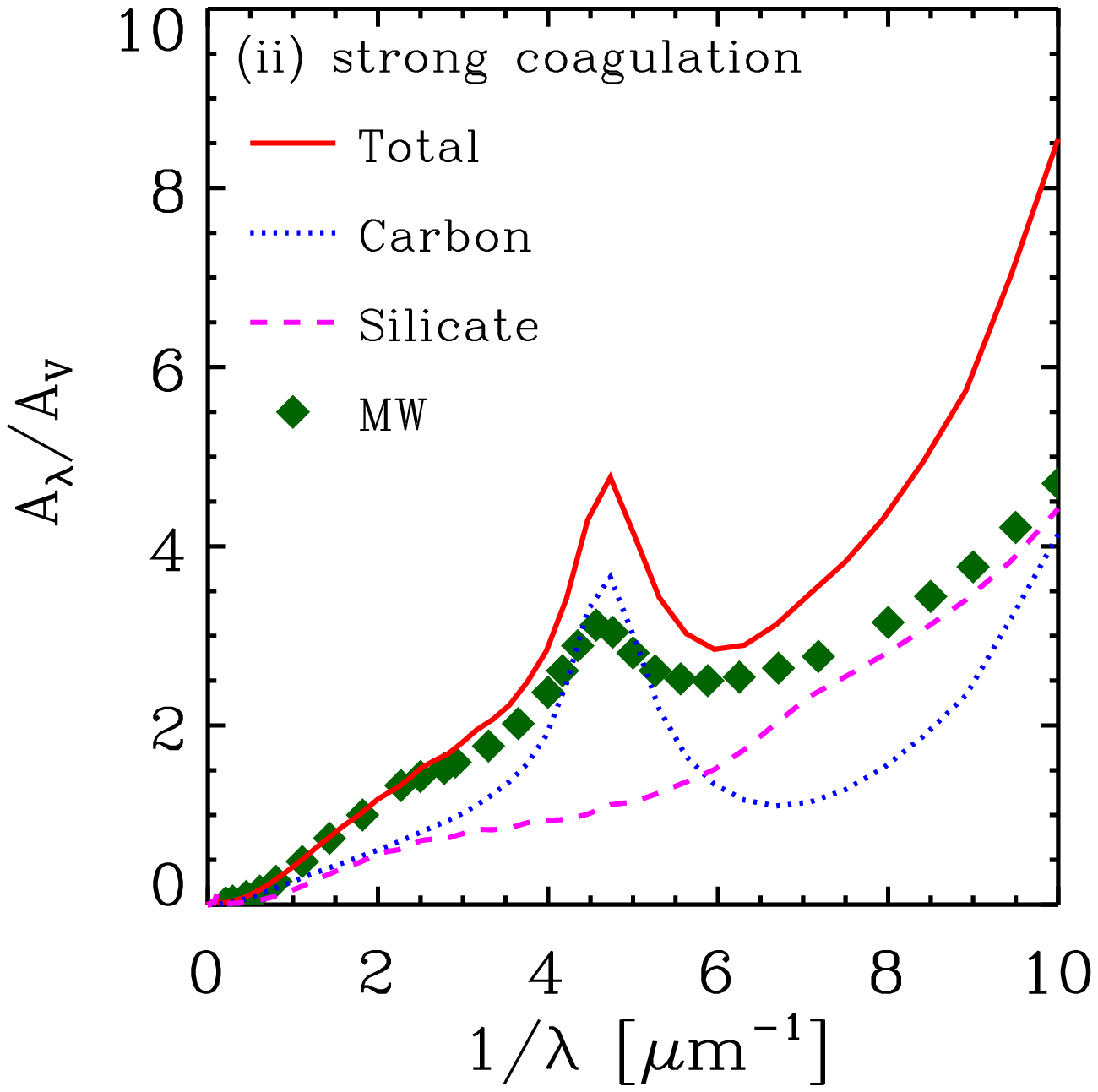}
\includegraphics[width=0.4\textwidth]{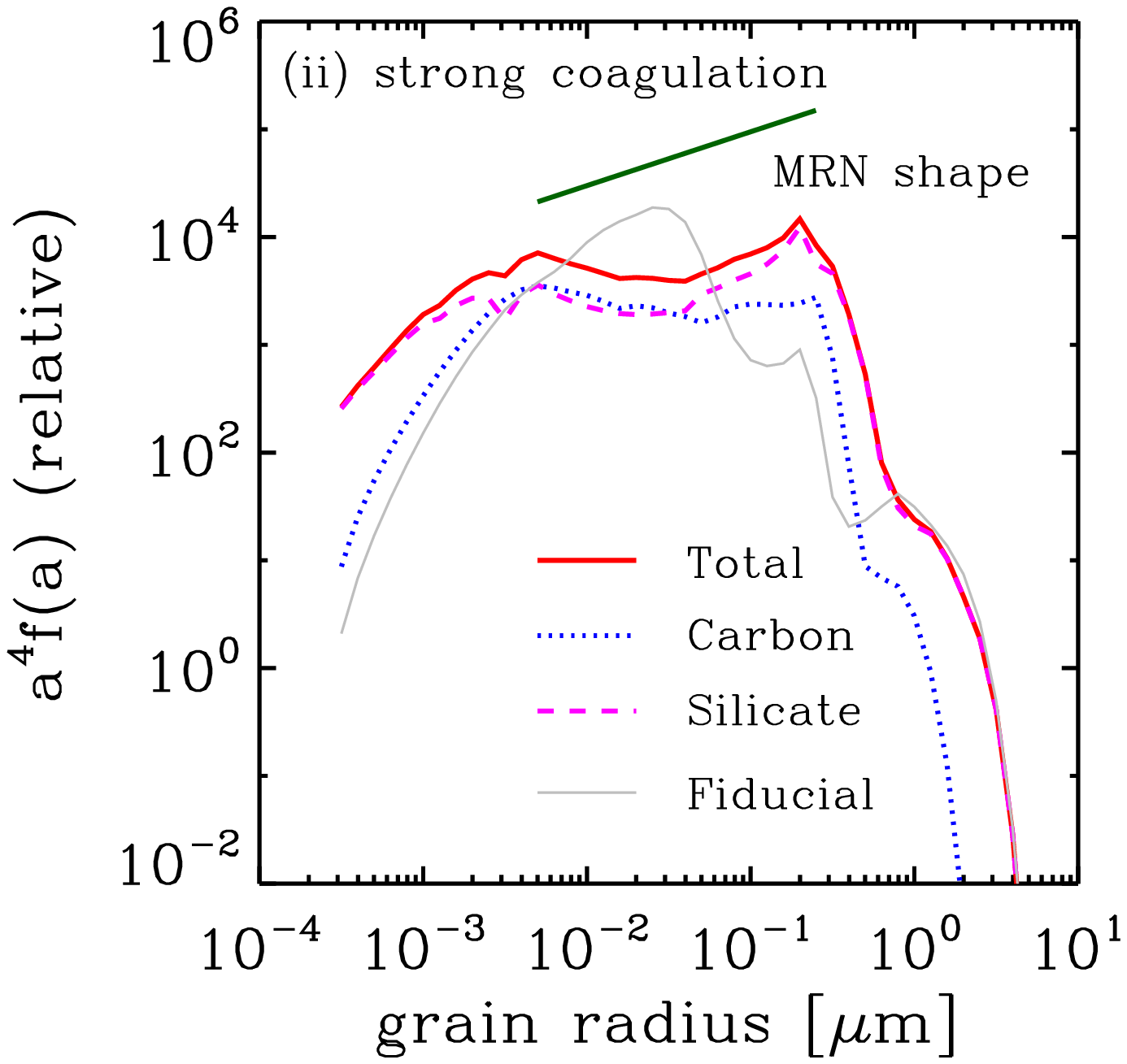}
\caption{Extinction curves (left panels) and grain size
 distributions (right panels) at galactic age $10$~Gyr for total grains
 (solid line), carbon dust (dotted line) and silicate dust (dashed
 line).
 From top to bottom, two panels represent for the cases (i) without 
  grain growth, and (ii) without coagulation threshold velocity.
For reference, we plot the result of grain size distribution using
  the same parameters in section \ref{subsec:coag} (thin solid line), and the MW
  extinction curve (filled diamond) \citep{whittet}.
We adopt $\tau_{\rm SF} = 5$~Gyr and $\eta_{\rm WNM} = \eta_{\rm CNM} = 0.5$.
}
\label{fig:discuss}
\end{figure*}

Figure~\ref{fig:discuss} shows the extinction curves and the grain size
distributions for carbon and silicate dust at the galactic age $10$~Gyr for
the (i) no grain growth model and (ii) strong coagulation model.
We adopt $\tau_{\rm SF} = 5$~Gyr, $\eta_{\rm WNM} = \eta_{\rm CNM} =
0.5$.
For comparison, we show the grain size distribution using the same
parameters as in
section \ref{subsec:coag} for a fiducial case.
In both panels, we find that the MW extinction curve is better
reproduced than the cases considered in Section \ref{sec:result}.
For the no grain growth model [panel (i)],
we observe that the calculated extinction curve is similar to that of
the Milky Way since the shape of the grain size distribution is similar
to the MRN distribution.
This is because when grain growth is not considered, the bump produced
at $a < 0.01\;\mu$m in the size distribution (see Fig.~\ref{fig:gsd}) cannot be formed.
Consequently, the grain size distribution approaches a power-law size
distribution with a power index $\sim -3.5$ because shattering and coagulation are dominant
processes to form the grain size distribution \citep[e.g.,][]{kobayashi10}.
However, if we do not consider the contribution of grain growth to the
evolution of the total dust mass in galaxies, it is hard to
reproduce the total dust mass of the Milky Way as has been pointed out
by e.g., \citet{zhukovska,draine09a,inoue11,asano13a}.
For example, \citet{inoue11} showed that in the case without grain
growth, the total dust amount in a galaxy is about ten times less than
that in the Milky Way even if dust grains are not destroyed by SN shocks.
Thus, by this scenario, we can obtain the extinction curve similar to
that of the Milky Way, but has difficulty in reproducing the total dust mass.

Next, we discuss the extinction curve for the case of strong coagulation
[panel (ii)].
We observe that the extinction curve has a slightly larger bump and a steeper
UV slope than the MW extinction curve, since the abundance of small grains ($a < 0.01\;\mu$m) is slightly
enhanced compared with the MRN size distribution.
However, we can find that the bump in the grain size distribution
disappears.
In this model, we remove the coagulation threshold, 
so the bump in the grain size distribution can shift toward
larger sizes than the fiducial case.
When the radii of grains reach at $\sim 0.2\;\mu$m, the grains
are shattered effectively, and consequently, the bump is vanished.
Furthermore, thanks to the supply of large grains by strong coagulation,
shattering is induced, and as a result, the small grains are supplied.
Hence, the amount of the small grains is larger than the fiducial case.
With such an interplay, the size distribution is expected to approach the power-law size
distribution with a power index $-3.5$ \citep{tanaka,kobayashi10}.
Consequently,
the extinction curve is nearer to the MW extinction curve than the cases
in Section~\ref{sec:result}.
In addition, unlike the case (i), the case (ii) may naturally account
for the evolutions of the total dust mass, grain size distribution
and extinction curve in a galaxy at the same time.
Note that power-law-like grain size distributions can also reproduce the
LMC and SMC extinction curves \citep{pei}.
Thus, we conclude that the strong interplay of shattering and coagulation is important to reproduce
extinction curves observed in nearby galaxies.

\subsection{Other possible grain species}

As we mentioned in Section \ref{sec:intro}, we do not consider PAHs
in this work.
PAHs are important in reproducing the SEDs at 
near- and mid- infrared wavelengths in galaxies \citep[e.g.,][]{lidra01}, and 
PAHs may contribute to the $2175$~\AA~bump in the MW
extinction curve \citep[e.g.,][]{WD01}.
However, what kind of grains dominate the bump at
$2175$~\AA~is still a matter of debate \citep[e.g.,][]{draine93,lidra01,draine03}.
Furthermore, although carbon-rich AGB stars
or shattering process in grain{--}grain collisions are considered as
possible sources of PAHs \citep[e.g.,][]{latter,jones96,seok13}, the main
formation mechanism of PAHs is still controversial.

We do not consider other carbonaceous dust, such as glassy carbon or
amorphous carbon.
\citet{nozawa13} discussed the effect of these species on the extinction curves, and
showed that they make the bump at $2175$~\AA~small. 
This means that the $2175$~\AA~bump which we calculated may be
overestimated.
Recently, \citet{jones13} have constructed a dust model considering
hydrogenated amorphous carbon without graphite, and the model reproduced
the extinction curve, IR extinction, IR-mm dust emission, and albedo of
the Milky Way.
In fact, their best fitting model also depends on the dust properties adopted.
However, the trend that the small carbonaceous grains produce the $2175$~\AA~bump and 
large grains show the flat extinction does not change.
Thus, the evolutionary trend of the shape of extinction curves
would not change significantly even if species other than graphite are
adopted as a representative carbonaceous
species.

In this paper, we consider the astronomical silicate as one of the main dust
species.
However, many observations have suggested that other species of cosmic silicate may
exist as iron-poor ones such as enstatite ($\mbox{MgSiO}_3$) and
forsterite ($\mbox{Mg}_2\mbox{SiO}_4$) \citep[e.g.,][]{draine03}.
In addition, many of the iron atoms are likely to
be locked up in pure Fe grains \citep{draine09b} and other Fe-bearing
grains such as magnetite ($\mbox{Fe}_3\mbox{O}_4$) and iron sulfide
(FeS) \citep[e.g.,][]{cowie}.
Recently, \citet{nozawa13} showed that the extinction curve produced by
Fe or $\mbox{Fe}_3\mbox{O}_4$ or FeS grains in addition to the
combination of graphite and $\mbox{Mg}_2\mbox{SiO}_4$ is similar to that led by the
combination of graphite and astronomical silicate, as long as most of
Mg, Si, and Fe atoms are locked up in these grains.
This indicates that the astronomical silicate and the
combination of other dust species have similar optical properties to each other.
Thus, even if we consider other species of silicate grains,
it is unlikely that this would largely modify a discrepancy
between the modelled and MW extinction curves.

\subsection{Remarks on extinction in high-$z$ galaxies}

In Section~\ref{sec:mw}, we compared our models with the MW
extinction curve, and investigated what the important physics that
reproduce the MW extinction curve is.
Here, we consider implication of our results for observed extinction
curves in high-$z$ galaxies.

Recently, the extinction curves in high-$z$ galaxies have been explored
by many authors \citep[e.g.,][]{maiolino,gallerani,hjorth}.
Indeed, from observations, \citet{gallerani} showed that the extinction curves in
high-$z$ ($z > 4$) QSOs tend to be flatter than those in nearby galaxies.
In addition, these flat extinction curves in high-$z$ galaxies are thought to
be dominated by dust grains ejected by SNe~II whose lifetime is short
(typically $10^{6-7}$~yr) \citep[e.g.,][]{maiolino,hirashita08}.
Our result that the extinction curves in the early stage of galaxy evolution
are relatively flat, is likely consistent with their results, although
our extinction curves may be too flat.
However, there is a suggestion that the large dust amount of the
dusty QSOs in high-$z$ Universe is regulated by grain growth
\citep[e.g.,][]{michalowski,kuo}.
This claim is strengthened by the fact that these dusty QSOs in high-$z$
Universe have metallicity $Z > Z_{\odot}$ \citep[e.g.,][]{juarez}.
As shown by our calculations, once grain growth occurs, the extinction curve
changes significantly.
Thus, as we tried in Section~\ref{sec:mw}, strong coagulation may be
worth considering also for high-$z$ QSOs.
However, we should note that QSOs with steep extinction curves may be
missed from the sample
because of the strong UV (optical in the observer's frame) extinction.

The rest-UV SEDs in high-$z$ galaxies ($z \ga 4$) have been investigated
by several authors \citep[e.g.,][]{bouwens,gonzalez}.
In particular, it is reported that UV colors in these galaxies become bluer with
increasing redshift \citep[e.g.,][]{bouwens}.
This fact is generally interpreted as little dust amount in galaxies.
However, the relation between extinction and UV color depends on the
extinction curve \citep{wilkins}.
In particular, the blue color may be explained with a flat extinction curve
shown by our results in the early stage of galaxy evolution.
Thus, there may be a significant amount of dust even in high-$z$ galaxies
with blue colors, which implies that the intrinsic luminosity may be
underestimated.

The strong $2175$~\AA~feature in star
forming galaxies at $1 < z < 2.5$ (the age of the Universe is $\sim
3${--}$6$~Gyr in a $\Lambda$CDM cosmological model with
$\Omega_{\Lambda} = 0.7$, $\Omega_{\rm M} = 0.3$, and $H_{0} = 70\;{\rm
km}~{\rm s}^{-1}~{\rm Mpc}^{-1}$) is reported by \citet{noll07}.
They found the robust evidence of the $2175$~\AA~bump in one-third of
their sample.
The existence of the $2175$~\AA~bump in galaxies at these redshifts
supports our results that the bump becomes strong in galaxies at $t
\sim \tau_{\rm shat} \sim 1\;(\tau_{\rm SF}/{\rm Gyr})^{1/2}$~Gyr with
$\tau_{\rm SF} \la 10$~Gyr.
Thus, shattering and grain growth may be effective in these galaxies,
and the variation of the extinction curve can be explained by the
difference of the galactic age and the star formation timescale.

\section{Conclusion}
\label{sec:conclusion}

We investigated the evolution of extinction curve in galaxies 
based on the evolution model of grain size distribution (A13). 
We took into account
various dust processes; dust formation by SNe~II and AGB stars, dust
reduction through astration, dust destruction by SN shocks in the ISM,
grain growth in the CNM, grain{--}grain collisions (shattering and
coagulation) in the WNM and CNM.

We found that the extinction curve is flat in the early stage of galaxy
evolution because of large grains ($a \ga 0.1\;\mu$m)
produced by stars.
As the galaxy evolution proceeds, shattering becomes effective, and the
number of small grains increases.
Further on, these small grains grow due to grain growth, forming a bump
at $a \sim 10^{-3}${--}$10^{-2}\;\mu$m in the grain size distribution.
Since the relatively small grains are dominant in grain size distribution, the extinction curve has a large bump at $1/\lambda
\sim 4.5\;\mu{\rm m}^{-1}$ and steep UV slope.
The timescale on which dust processes in the ISM (grain growth, shattering and
coagulation) begin to control the grain size distribution, is
estimated as $t \sim 1 (\tau_{\rm SF}/{\rm Gyr})^{1/2}$~Gyr.
After coagulation occurs effectively, the extinction curves become
flatter but still tend to be steeper than the MW extinction curve.
We also found that for reproducing the MW extinction curve, 
it may be important to consider a larger contribution of coagulation than
that we have assumed.
This means that the strong interplay between shattering and coagulation
induced by strong coagulation could be essential to reproduce the MW extinction curve.
We conclude that the extinction curves in galaxies drastically
changes with galactic age because of the evolution of grain size
distribution.

\section*{Acknowledgments}

We thank the anonymous referee for her/his helpful comments which
improved the quality and clarity of this paper.
We are grateful to Takashi Kozasa, Anthony P. Jones, Daisuke Yamasawa, and Asao Habe for fruitful discussions.
We thank H. Kobayashi for helpful discussions on the
process of the grain{--}grain collisions and the evolution of the grain size distribution.
RSA has been supported from the Grant-in-Aid
for JSPS Research under Grant No. 23-5514.
TTT has been supported by the Strategic
Young Researcher Overseas Visits Program for Accelerating
Brain Circulation No. R2405 and the Grant-in-Aid for the Scientific 
Research Fund (20740105, 23340046) commissioned by the MEXT.
HH is supported by NSC grant 102-2119-M-001-006-MY3.
TN has been supported by World Premier International Research Center
Initiative (WPI Initiative), MEXT, Japan, and by the Grant-in-Aid for
Scientific Research of the JSPS (22684004, 23224004).

%\appendix

\bsp

\label{lastpage}

\end{document}